# Direct Experimental Observation of the Gas Filamentation Effect using a Two-bunch X-ray FEL Beam

Authors

**Y. Feng\*, D. W. Schafer, S. Song, Y. Sun, D. Zhu, J. Krzywinski, A. Robert, J. Wu, and F.-J. Decker**

SLAC National Accelerator Laboratory, 2575 Sand Hill Road, Menlo Park, California, 94025, USA

Correspondence email: yfeng@slac.stanford.edu

**Synopsis**  Direct experimental observation of the gas filamentation effect by performing an X-ray pump-X-ray probe measurement using a Two-bunch X-ray FEL beam.

**Abstract**  We report the experimental observation of the filamentation effect in gas devices designed for X-ray Free-electron Laser's. The measurements were carried out at the Linac Coherent Light Source on the X-ray Correlation Spectroscopy (XCS) instrument using a Two-bunch FEL beam at 6.5 keV with 122.5 ns separation passing through an Argon gas cell. The relative intensities of the two pulses of the Two-bunch beam were measured, after and before the gas cell, from the X-ray scattering off thin targets by using fast diodes with sufficient temporal resolution. It was found that the after-to-before ratio of the intensities of the second pulse was consistently higher than that of the first pulse, revealing lower effective attenuation of the gas cell due to the heating and subsequent gas density reduction in the beam path by the first pulse. This measurement is important in guiding the design and/or mitigating the adverse effect in gas devices for high repetition-rate FEL's such as the LCLS-II and the European XFEL or other future high repetition-rate upgrade to existing FEL facilities.

## 1. Introduction

The phase II project of the Linac Coherent Light Source (LCLS) (Emma *et al*., 2010) will seek to upgrade the LCLS conventional accelerator to a superconducting Linac, similar to that of the upcoming European XFEL, to generate X-ray Free-Electron Laser (FEL) beams at a much higher repetition rate up to 1 MHz. The high repetition-rate operation of the European XFEL and the LCLS-II is expected to provide additional transformative capabilities to those already offered by the low repetition-rate facilities that are either currently in user operation or being commissioned, including FLASH (Ackermann *et al*. 2007) and its upgrade FLASH-II, LCLS, SACLA (Ishikawa, *et al*., 2012), FERMI@Elettra (Allaria *et al*., 2012), PAL-XFEL (Kang, Kim, and Ko, 2013) and SwissFEL (Ganter *et al*., 2010). At the same time, these new capabilities also bring about new challenges in conceiving, designing and implementing high repetition-rate compatible X-ray diagnostic, optics, beam regulation, and safety devices, which have been proven to be very important to having helped fulfill FEL's great scientific potentials in the frontier research of physics, chemistry, life science, material, energy and Earth sciences (Young *et al*., 2010; Glover *et al*., 2012; Fuchs *et al*., 2015; Minitti *et al*., 2015; Chapman *et al*., 2011; Seibert *et al*., 2011; Gerber *et al*., 2015; Harmand *et al*., 2015). The ever so importance of X-ray diagnostics for an FEL stems from the stochastic nature of its lasing mechanism, i.e., the Self-Amplified Spontaneous Emission (SASE) process (Kondratenko and Saldin, 1979; Bonifacio, Pellegrini, and Narducci, 1984), which gives rise to shot-to-shot fluctuations in all beam properties, requiring single-shot diagnostics in pulse energy, timing, spectrum, and etc. The successful constructions of these diagnostics devices (Feng *et al*., 2011; Bionta *et al*., 2011; Zhu *et al*., 2012) have allowed users to gain sensitivity especially in X-ray pump probe experiments (Chollet *et al*., 2015) which are otherwise difficult to discern, as well as enabled FEL operators for not only tuning up the FEL performance but also developing new operating modes (Amann *et al*., 2012; Marinelli *et al*., 2013; Lutman *et al*., 2014).

At soft X-rays energies especially, FEL diagnostics and components are often limited to those based on a gas medium to circumvent single-shot damage risks posed by the enormous peak power of an FEL beam. For example, gas intensity monitors and attenuators have been implemented at various facilities including FLASH (Richter *et al*., 2003; Hahn and Tiedtke, 2007) and FALSH-II, and LCLS (Hau-Riege *et. al*., 2008; Ryutov *et al*., 2009) and are being planned for LCLS-II. The LCLS has the highest operating frequency of all low repetition-rate facilities, in user operation now or otherwise, at 120 Hz, leaving a time $\Delta T$ of ~ 8.3 ms for the gas system to return to its starting density after the passage of each pulse. A fast energy dissipation has always been implicitly assumed for longer $\Delta T$, but will start to break down as $\Delta T$ will be reduced by more than four orders of magnitude for the European XFEL or LCLS-II to less or equal to 1 µs. The impact of the very short $\Delta T$ on the performance of these gas diagnostics and devices has been studied extensively using thermodynamic and hydrodynamic simulations (Feng *et al*., 2016a; Feng *et al*., 2015a; Feng *et al*., 2015b; Yang *et al*., 2017, Feng *et al*., 2017), revealing nonlinear effects in the precision of intensity measurements for gas monitors as well as the effective attenuation for gas attenuators, stemming from the density depression or filamentation phenomenon induced by the energy deposition of the preceding pulses on trailing ones. Parallel efforts have been put forth to experimentally confirm and quantify the filamentation effect, first by using an optical pump and optical probe technique similar to what was first performed with an ultrafast optical laser (Cheng *et al*. 2013), and then an X-ray pump and optical probe measurement (Galtier and Schafer 2017).

In this report, we present the first direct experimental observation of the gas filamentation effect by an ultrafast X-ray FEL beam using an X-ray pump and X-ray probe technique. The measurements were carried out at the LCLS on the X-ray Correlation Spectroscopy (XCS) instrument (Alonso-Mori *et al*., 2015) at an X-ray energy of 6.5 keV. Using a special operating mode of the LCLS, a Two-bunch FEL beam having two similar intensity femtosecond pulses, which were spatially collinear but separated by over 100 ns in time, was generated. It was sent through the Argon-filled gas cell, and the intensities of the two pulses were measured independently, before and after the gas cell. It was found that the transmission of the second pulse was consistently higher than that of the first pulse, revealing a lower effective attenuation consistent with the gas filamentation effect predicted by thermodynamic and hydrodynamic calculations (Feng *et al*., 2016a; Feng *et al*., 2015a; Feng *et al*., 2015b; Yang *et al*., 2017, Feng *et al*., 2017).

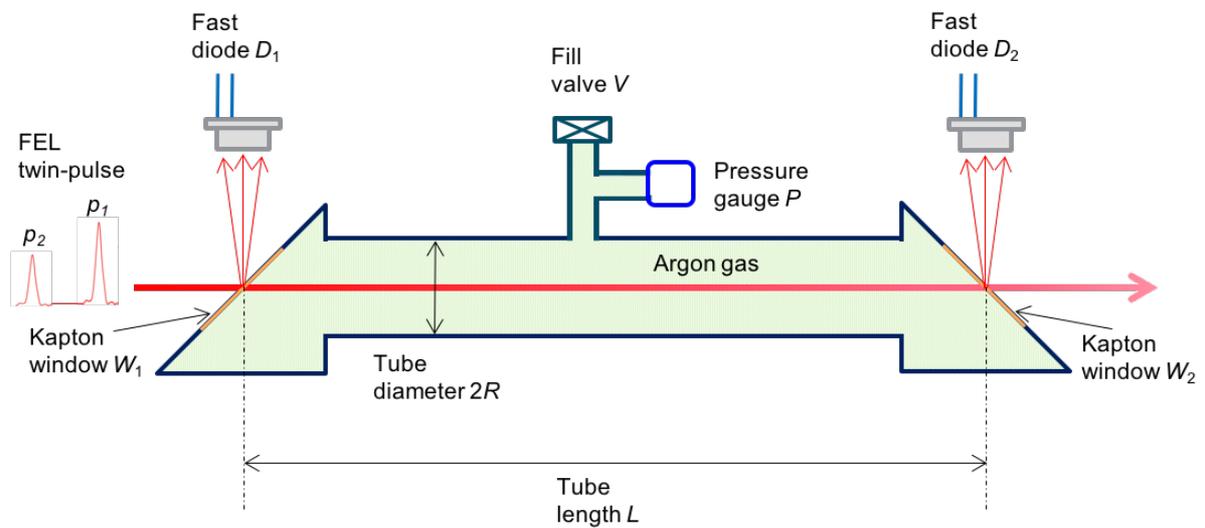

**Figure 1** Schematic of the experimental setup, consisting of a gas cell of diameter of $2R$ and effective length $L$, filled with Argon gas to a pressure $P$. The cell was capped off by two 50 µm thick Kapton windows $W_1$ and $W_2$ each inclined to 45° to the direction of an X-ray Two-bunch FEL beam coming from the left and being attenuated. The X-ray scatterings off the Kapton windows were measured as the intensities of $p_1$ and $p_2$ by two fast diodes $D_1$ and $D_2$ positioned perpendicular to the beam and each connected to a different channel of a fast digitizer.

## 2. Experimental Setup

To perform a direct X-ray pump and X-ray probe of the gas filamentation phenomenon at LCLS before the arrival of LCLS-II, or the availability of a hard X-ray split-and-delay device, the two X-ray pulses were generated using an accelerator-based approach. In a specially developed operating procedure, or the so-called "Two-bunch operating mode", two spatially collinear and of similar intensity pulses $p_1$ and $p_2$ could be produced with $p_2$ delayed by a time $\Delta t$ of a multiple of 0.35 ns (Decker, *et al*., 2010). For reasons that will become clear in later discussion, an optimal delay of order 1 µs would have been ideal for the current experiment, but rather a delay of only 122.5 ns was used due to a limitation of the Two-bunch mode. The average intensity of the first pulse $p_1$ was estimated to be about 1 mJ entering into the gas cell and was purposely tuned to be greater than that of $p_2$ to maximize the pump and yet to retain sufficient signal-to-noise ratio of for the probe.

The Two-bunch $p_1$ and $p_2$ FEL beam was directed through an Argon filled gas cell depicted schematically in Figure 1, and being attenuated. The gas cell had a diameter $2R = 22.1$ mm and was sealed off by two 50 μm thick Kapton windows $W_1$ and $W_2$, each inclined to 45° to the direction of the FEL beam and separated by a nominal distance of $L = 165$ mm. The gas pressure $P$ was measured via an MKS Baratron gauge (Model No. 722B23TGA2FA), with a reading precision of 0.5%, and was varied from $P = 0$ to $P = 193$ Torr. To minimize gas flows unrelated to the underlying physics under investigation, the gas cell was valved off, but a small leak in the Kapton window developed over time and did lead to a slow upward creeping of the gas pressure on a time scale of many tens of minutes, requiring making pressure adjustment in the data analysis.

The total X-ray scattering, including both the coherent Thomson, incoherent Compton ad other processes, off the Kapton windows was measured via two Hamamatsu diodes (Model No. MSM Photodetector G4176-03), $D_1$ and $D_2$, positioned perpendicular to the beam. The sides of the diodes were shielded from scatterings, but no metal covers were used to block any optical light. The rise time of the diodes was estimated to be better than 1 ns, and each diode was connected directly to a different input channel (via a 50 Ω coupling resistor) of a fast digitizer (Aquiris 10-bit High-Speed cPCI digitizer, Model No. U1065A). The sampling frequency was set to 0.25 ns and there was a total of 4000 sampling channels, generating a 1 μs range of digitization. As such, the contribution in the dark noise of the digitizer by ambient light was completely negligible, although any concurrent optical emission upon the passage of the X-ray pulses within the digitization window cannot be separated from the signal. The scattering intensities of $p_1$ and $p_2$ were used as measurements of the pulse intensity $I_{1i}$ and $I_{2i}$, where $i = 1$ for the upstream measurement, and $i = 2$ for downstream measurement.

The nominal photon energy of the Two-bunch FEL beam was 6.5 keV, with each pulse lased independently and retained typical SASE characteristics, including the large fluctuations in intensity. The two pulses could be steered independently, and care was taken to ensure their spatial overlap over the length of the gas cell by about 80% in size using a downstream high resolution beam imager. Note that a complete overlap is in principle not achievable due to the SASE nature, which produces approximately 10% spatial jitter pulse-to-pulse. The time delay was fixed at 122.5 ns, the maximum achievable with the LCLS accelerator configuration at the time of the measurement. As such, the rise time of the diodes (< 1 ns) was sufficient to resolve the twin pulses, making the data analysis very straightforward without having to be concerned about temporally overlapping signals.

The Two-bunch FEL beam was about 1 mm in size in the XCS hutch before focusing, and the formation of the filament upon passage of an FEL beam takes place over the time scale set by the speed of the shock wave $v_{\text{shock}}$ (Yang *et al*., 2017) at 200 m/s, which is approximately on same scale as the speed of sound in an ideal gas ($v_{\text{sound}} \sim 300$ m/s). As such, the size of the filament only reaches about 24 µm after 122.5 ns, much smaller than the beam size of 1 mm, making it much more difficult to observe. By focusing the beam down to about 100 µm on average over the length of the cell, clear evidence of enhanced transmission of the second pulse as induced by the first pulse was found, in qualitative agreement with the simulation results.

## 3. Data Analysis and Results

### 3.1. Transmissions at zero attenuation

The average traces of the pulses $p_1$ and $p_2$ from the digitizer of the Two-bunch pulses at $P = 0$ Torr are shown in Figure 2, with those of measured by diode $D_2$ somewhat larger and at the same time narrower by about 16±1 % thus summing up to the same integrated intensity. This is due to the slight difference in the response time of the two diodes not being exactly the same. The average was done over more than 32,000 pulses. There were two contributions to the background, one of random nature and the other periodic and alternating exactly at first subharmonic of the sampling frequency of 0.25 ns of the digitizer, indicating a possible digital artifact. The second periodic contribution was simply removed by applying a filter at half of the sampling frequency, thus is not visible in Figure 2.

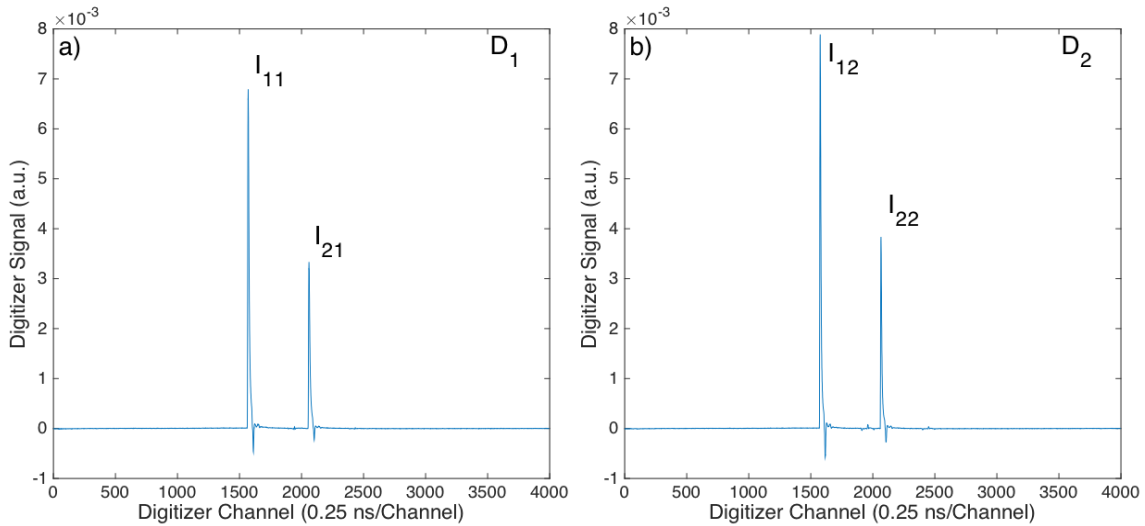

**Figure 2** The average digitizer traces of the Two-bunch beam measured by a) the upstream diode $D_1$ and b) the downstream diode $D_2$ at an Argon pressure $P = 0$ Torr. The digitizer was set to sample at 0.25 ns per channel. The integrated intensities of the first pulse $p_1$ are denoted as $I_{11}$ and $I_{12}$ on $D_1$ and $D_2$, respectively, and those of the second pulse $p_2$, as $I_{21}$ and $I_{22}$ on $D_1$ and $D_2$, respectively.

The first pulse $p_1$ and second pulse $p_2$ were separated by exactly 122.5 ns or 490 sampling channels, and the rise time of the upstream diode $D_1$ was about 2 ns, whereas for the downstream diode $D_2$, it was about 1.75 ns, consistent with the 16% difference in their respective amplitudes. The amplitude of the undershoots in traces was also proportional to the main peak, and was caused by a slight impedance mismatch or ringing. The relative intensities of the first pulse $I_{1i}$ and second pulse $I_{2i}$ on different diodes $D_i$, for $i = 1$ or 2, were obtained by integrating the digitizer traces over the main peak (not including the undershoot), and the average transmission coefficients of the two pulses $T_j = I_{j2}/I_{j1}$, for $j = 1$ and 2, were then calculated, obtaining $<T_1> = 1.01\pm0.02$ and $<T_2> = 0.99\pm0.02$, respectively. The near equality of $<T_1>$ and $<T_2>$ within the experimental uncertainty was expected as there was no attenuation at $P = 0$ Torr.

The Hamamatsu G4176-03 diode was chosen to achieve sufficient time resolution, but at the expense of a small dynamic range for detecting hard X-rays. At 6.5 keV, a large number of electron-hole pairs, approximately 1800, are created by a single photon. To avoid saturation, the number of scattered photons into the diodes were purposely limited to a few tens by adjusting their distances to the Kapton windows. As such, large uncertainties were observed in the single-pulse measurement of $I_{ji}$ for $i, j = 1$ and 2, as shown in Figure 3a) and Figure 3b) and were believed to be of Poisson nature. At a given measured relative intensity $I_{11}$, or $I_{21}$, there is a very wide range of values for $I_{12}$ or $I_{22}$, with the standard deviation reaching as much as 25%. However, the correlations such as in Figure 3 still provided consistent transmission measurements with those obtained from the average traces in Figure 2, as given by the slope of a linear fit, at $<T_1> = 0.975\pm0.003$ and $<T_2> = 0.976\pm0.001$ for the first and second pulses, respectively.

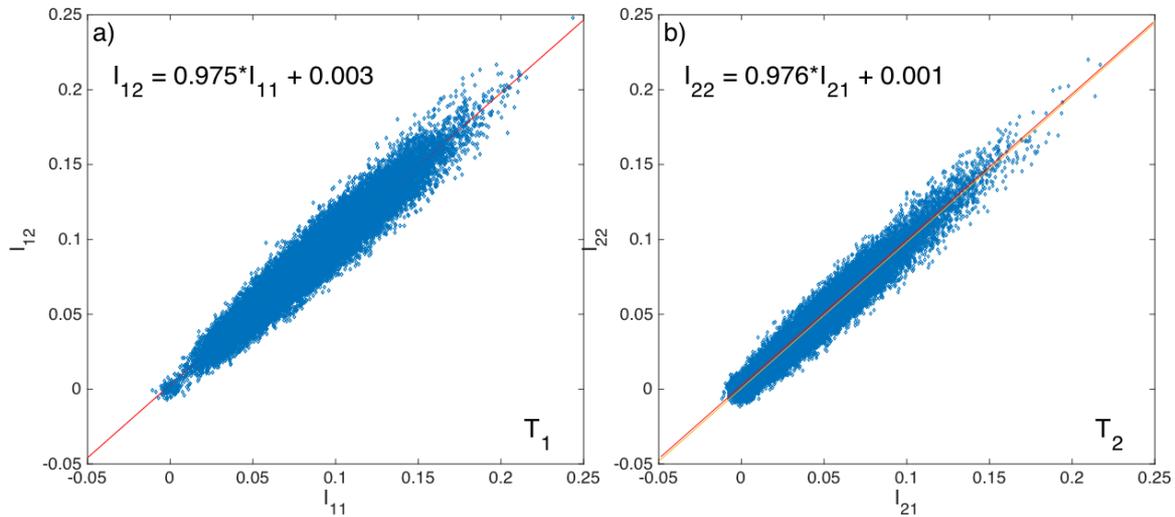

**Figure 3** The correlations between the intensities measured by the upstream and downstream diodes of a) the first pulse, and b) the second pulse at Argon gas pressure $P = 0$ Torr. The slope is a measure of the transmission, amounting to $<T_1> = 97.5\%$ for the first pulse and nearly identical $<T_2> = 97.6\%$ for the second pulse as expected. The very slight difference from unity is considered as a systematic error of the detection scheme. The red line for the first pulse was overlaid in b) to stress the similarities between the transmissions of the two pulses.

### 3.2. Transmissions at finite attenuation

Additional measurements similar to those at zero attenuation were repeated at finite attenuations. The average traces of the pulses $p_1$ and $p_2$ from the digitizer of the Two-bunch pulses at an effective Argon pressure of $P = 109$ Torr are shown in Figure 4. During the measurement at $P = 0$ Torr, the FEL beam was focused from about 1 mm in size to 100 μm at the location of the gas cell to gain sensitivity. The focused beam was starting to drill a pinhole into the Kapton window, creating a tiny leak. As such, the reading on the pressure gauge at 98.8 Torr was lower than the actual transmission would indicate. Again, the systematic periodic background was first removed. Similar to the zero pressure case, the trace average was done over more than 35,000 pulses. The relative intensities on the downstream diode were lower as expected than those on the upstream but by different amounts in relative terms, indicating different effective transmissions for the two pulses.

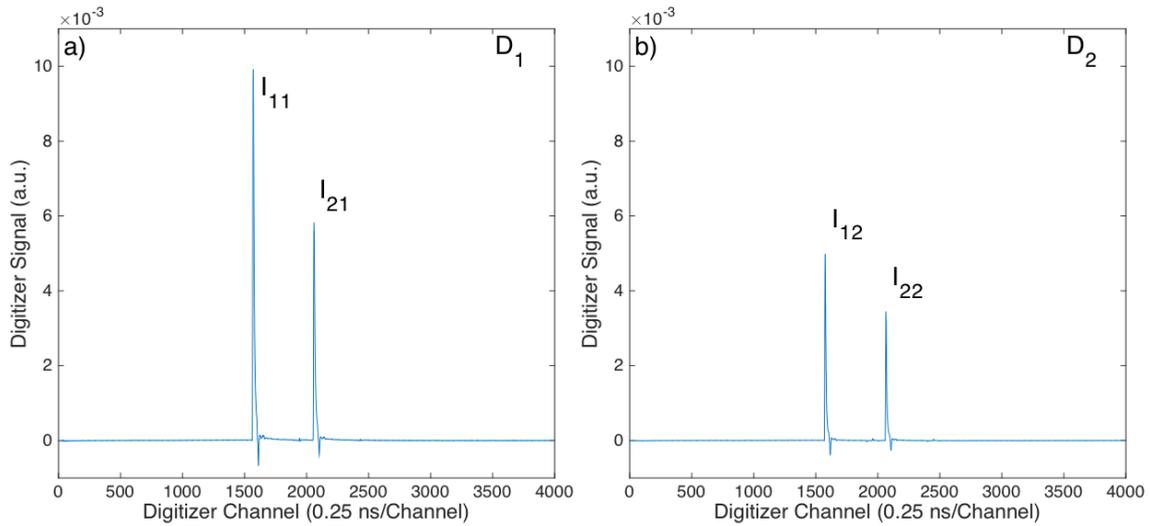

**Figure 4**  The average digitizer traces of the Two-bunch beam measured by a) the upstream diode $D_1$ and b) the downstream diode $D_2$ at an effective Argon pressure $P = 109$ Torr. The various intensities $I_{ij}$ are defined previously in Figure 2.

The different effective transmissions for the two pulses are best shown by the linear fits to the intensity correlation plots between the before- and after gas cell measurement in Figure 5, equalling to $<T_1> = 0.427$ for the first pulse, and a higher value $<T_2> = 0.498$ for the second pulse. This amounts to a 17% enhancement in the transmission of the second pulse induced by the first pulse. This difference is further elucidated by overlaying the calculated transmissions in Figure 6a) sorted by the intensity of the probe pulse $I_{11}$, and the transmission histograms in Figure 6b), both giving an average of $<T_1> = 43\%$ and $<T_2> = 50\%$ for the first and second pulse, respectively. The large uncertainty in the measurements due to the Poisson noise was again apparent in both figures, with the second pulse having roughly 50% greater noise, consistent with its reduced intensity at 43%, which should scale like $1/\sqrt{(0.43)} \sim 1.52$.

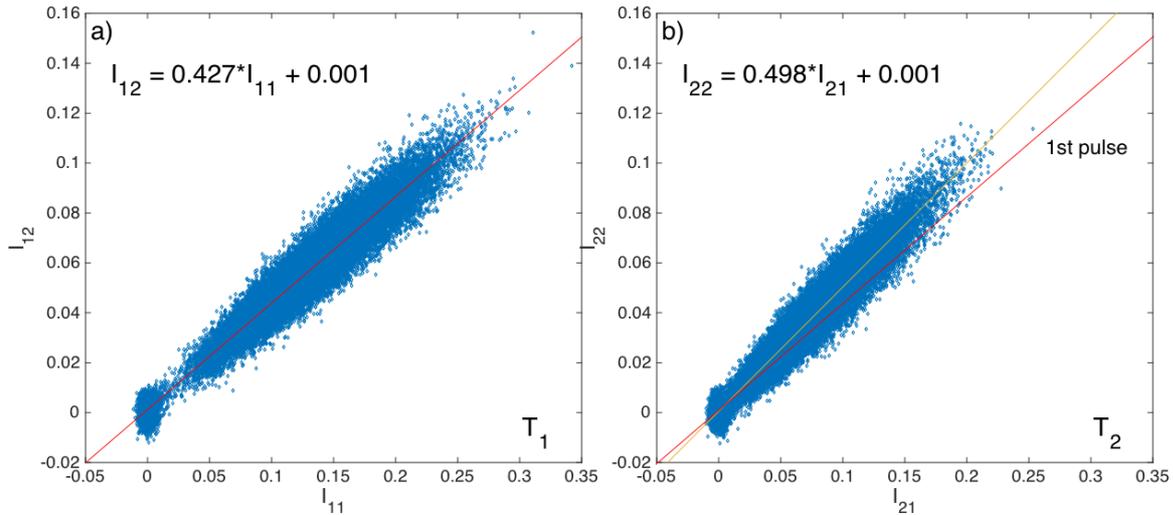

**Figure 5** The correlations between the intensities measured by the upstream and downstream diodes of a) the first pulse, and b) the second pulse at an equivalent Argon gas pressure $P$ = 109 Torr. The slope is a measure of the transmission, amounting to 42.7% for the first pulse and an enhanced 49.8% for the second pulse.

The enhanced transmission from $<T_1>$ = 43% to $<T_2>$ = 50% is much smaller than the $<T_2>$ = 75% value calculated based on the methodology developed in the previous thermodynamic studies (Feng, *et al*., 2015a) and the experimental geometries. There is a main reason for this discrepancy. The delay time $\Delta t$ at 122.5 ns was too short so the filament was not fully developed, or the density depression was not completely formed according to the recent hydrodynamic studies (Yang, *et al*., 2017). If using the sound speed of 200 m/s obtained in the hydrodynamic simulation and the focused beam size of 100 µm, the largest transmission would happen at approximately 500 ns. If simply extrapolate the development of the filament in time linearly, one would expect an enhanced transmission at $<T_2>$ = (75% - 43%)·122.5/500 + 43% = 51%, in excellent agreement with the measurement. This consistency also validates the time-dependent thermodynamic simulations (Feng, *et al*., 2015a) for time scales beyond the fully opening of the filament, which depends on the beam size and sound speed. For shorter time scales down to that of the thermalization (Feng, *et al*., 2016), hydrodynamic simulations are required to predict the exact transmission. Furthermore, the current experimental measurement also suggests that the main energy transfer mechanism is still thermodynamic, and any radiative losses or other processes would be secondary.

## 4. Summary

We have observed experimentally the filamentation effect in a gas cell induced by ultrashort X-ray FEL beam using a Two-bunch scheme. The observed enhanced transmission of the delayed second pulse in the wake of the first pulse was in good quantitative agreement with the estimate, which was based on thermodynamic simulations and proper extrapolation of the hydrodynamic simulation results. No enhancement was found as expected when the cell was not filled with an attenuating gas. Our finding is important in guiding the design and/or possibly mitigating the adverse effect in gas devices for high repetition-rate FEL's such as the LCLS-II, the European XFEL, or other future high repetition-rate upgrade to the existing FEL facilities, and have broader implications not only on gas-based applications, but also on liquid and solid based devices and experiments.

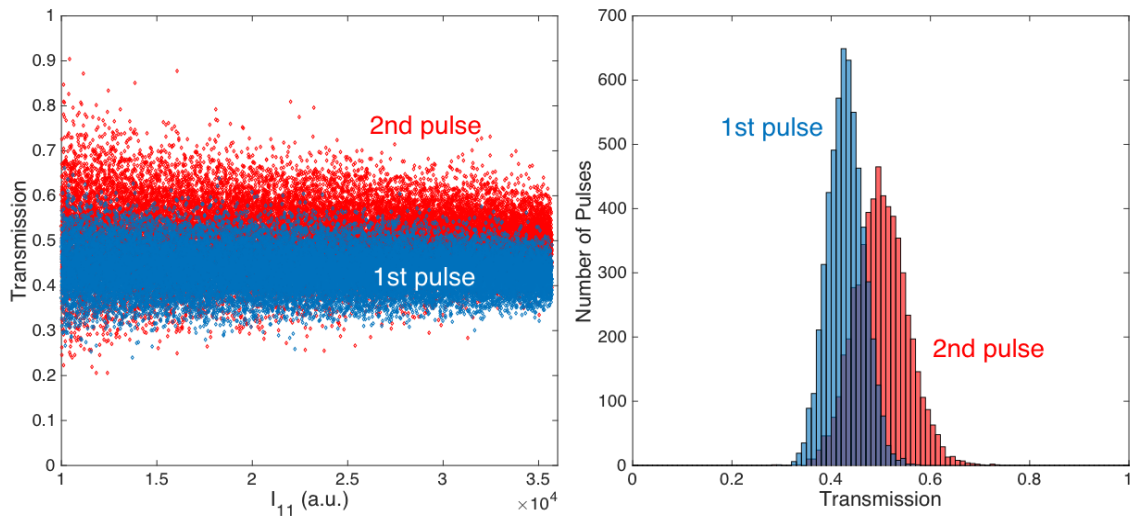

**Figure 6** a) The transmission of the first and second pulse sorted based on the intensity of the first or probe pulse $I_{11}$, at an effective Argon gas pressure $P$ = 109 Torr, with an average of $<T_1>$ = 43% and $<T_2>$ = 50% for the first and second pulse, respectively; b) The histogram of the transmission of the first and second pulse; The average transmissions are again $<T_1>$ = 43% and $<T_2>$ = 50% for the first and second pulse, respectively.

**Acknowledgements** Use of the Linac Coherent Light Source (LCLS), SLAC National Accelerator Laboratory, is supported by the U.S. Department of Energy, Office of Science, Office of Basic Energy Sciences under Contract No. DE-AC02-76SF00515. The authors also like to thank T. Raubenheimer, M. Rowen, E. Ortiz, E. Galtier, and Z. Huang for useful discussions and the LCLS XCS technical staff for their support of the experimental effort.

**References**

Ackermann, W., Asova, G., Ayvazyan, V., Azima, A., Baboi, N., Bähr, J., Balandin, V., Beutner, B., Brandt, A., Bolzmann, A. *et al.* (2007). *Nat. Photon.,* **1**(6), 336-342.

Allaria, E., Appio, R., Badano, L., Barletta, W., Bassanese, S., Biedron, S., Borga, A., Busetto, E., Castronovo, D., Cinquegrana, P. *et al.* (2012). *Nat. Photon.,* **6**(10), 699-704.

Alonso-Mori, R., Caronna, C., Chollet, M., Curtis, R., Damiani, D. S., Defever, J., Feng, Y., Flath, D. L., Glownia, J. M., Lee, S. *et al.* (2015). *J. Synchrotron Rad.*, **22**, 508-513.

Amann, J., Berg, W., Blank, V., Decker, F.-J., Ding, Y., Emma, P., Feng, Y., Frisch, J., Fritz, D., Hastings, J. *et al.* (2012). *Nat. Photon.,* **6**(10), 693-698.

Bionta, M. R., Lemke, H., Cryan, J., Glownia, J., Bostedt, C., Cammarata, M., Castagna, J.-C., Ding, Y., Fritz, D., Fry, A. *et al.* (2011). *Opt. Express,* **19**(22), 21855-21865.

Bonifacio, R., Pellegrini, C. & Narducci, L. M. (1984). *Opt. Commun.,* **50**(6), 373-378.

Chapman, H. N., Fromme, P., Barty, A., White, T. A., Kirian, R. A., Aquila, A., Hunter, M. S., Schulz, J., DePonte, D. P., Weierstall, U. *et al.* (2011). *Nature,* **470**(7332), 73-77.

Cheng, Y.-H., Wahlstrand, J. K., Jhajj, N. & Milchberg, H. (2013). *Opt. Express,* **21**(4), 4740-4751.

Chollet, M., Alonso-Mori, R., Cammarata, M., Damiani, D., Defever, J., Delor, J. T., Feng, Y., Glownia, J. M., Langton, J. B., Nelson, S. *et al.* (2015). *J. Synch. Rad.,* **22**(3), 503-507.

Emma, P., Akre, R., Arthur, J., Bionta, R., Bostedt, C., Bozek, J., Brachmann, A., Bucksbaum, P., Coffee, R., Decker, F.-J. *et al.* (2010). *Nat. Photonics*, **4**(9), 641-647.

Decker, F.-J., Akre, R., Brachmann, A., Ding, Y., Dowell, D., Emma, P. J., Fisher, A., Frisch, J., Gilevich, S., Hering, Ph. *et al.* (2010). *Proc. of FEL*, WEPB33.

Feng, Y., Feldkamp, J. M., Fritz, D. M., Cammarata, M., Aymeric, R., Caronna, C., Lemke, H. T., Zhu, D., Lee, S., Boutet, S. *et al.* (2011). *Proc. SPIE*, **8104**, 81040Q.

Feng, Y., Krzywinski, J., Schafer, D. W., Ortiz, E., Rowen, M. & Raubenheimer, T. O. (2015). *Proc. SPIE*, **9589**, 958910.

Feng, Y., Krzywinski, J., Campell, M., Schafer, D., Ortiz, E., Rowen, M. & Raubenheimer, T. O. (2015). *Proc. of FEL,* TUP026.

Feng, Y., Krzywinski, J., Schafer, D. W., Ortiz, E., Rowen, M. & Raubenheimer, T. O. (2016). *J. Synchrotron Rad.,* **23**(1) 21.

Feng, Y. and Raubenheimer, T.O. (2017). *Proc. of SPIE*, in press.

Fuchs, M., Trigo, M., Chen, J., Ghimire, S., Shwartz, S., Kozina, M., Jiang, M., Henighan, T., Bray, C., Ndabashimiye, G. *et al.* (2015). *Nat. Phys.,* **11**(11), 964-970.

Galtier, E. and Schafer, D. (2017). Unpublished.

Ganter, R., *et al*. (2010). *SwissFEL-Conceptual Design Report*, Technical Report, Paul Scherrer Institute, Villigen, Switzerland.

Gerber, S., Jang, H., Nojiri, H., Matsuzawa, S., Yasumura, H., Bonn, D., Liang, R., Hardy, W., Islam, Z., Mehta, A. *et al.* (2015). *Science,* **350**(6263), 949-952.

Glover, T., Fritz, D., Cammarata, M., Allison, T., Coh, S., Feldkamp, J., Lemke, H., Zhu, D., Feng, Y., Coffee, R. *et al.* (2012). *Nature,* **488** (7413), 603-608.

Hahn, U., Tiedtke, K., Choi, J.-Y. & Rah, S. (2007). *AIP Conf. Proc.* **879**, 276.

Harmand, M., Ravasio, A., Mazevet, S., Bouchet, J., Denoeud, A., Dorchies, F., Feng, Y., Fourment, C., Galtier, E., Gaudin, J. *et al.* (2015). *Phy. Rev. B*, **92**(2), 024108.

Hau-Riege, S., Bionta, R., Ryutov, D. & Krzywinski, J. (2008). *J. Appl. Phys.,* **103**(5), 053306.

Ishikawa, T., Aoyagi, H., Asaka, T., Asano, Y., Azumi, N., Bizen, T., Ego, H., Fukami, K., Fukui, T., Furukawa, Y. *et al.* (2012). *Nat. Photon.,* **6**(8), 540-544.

Kang, H.-S., Kim, K.-W., and Ko, I. S. (2013). *International Particle Accelerator Conference 2013*, 2074, Shanghai, China.

Kondratenko, A. & Saldin, E. (1979*). DOKLADY AKADEMII NAUK SSSR,* **249**(4), 843-847.

Lutman, A., Decker, F.-J., Arthur, J., Chollet, M., Feng, Y., Hastings, J., Huang, Z., Lemke, H., Nuhn, H.-D., Marinelli, A. *et al.* (2014). *Phy. Rev. Lett.,* **113**(25), 254801.

Marinelli, A., Lutman, A. A., Wu, J., Ding, Y., Krzywinski, J., Nuhn, H.-D., Feng, Y., Coffee, R. N., and Pellegrini, C. (2013). *Phys. Rev. Lett.*, **111**, 134801.

Minitti, M., Budarz, J., Kirrander, A., Robinson, J., Ratner, D., Lane, T., Zhu, D., Glownia, J., Kozina, M., Lemke, H. *et al.* (2015). *Phys. Rev. Lett.,* **114**(25), 255501.

Ryutov, D. D. *et al.* (2009). *Technical Report*, **LLNL-TR-421318**, Lawrence Livermore National Laboratory, Livermore, CA, USA.

Richter, M., Gottwald, A., Kroth, U., Sorokin, A., Bobashev, S., Shmaenok, L., Feldhaus, J., Gerth, C., Steeg, B., Tiedtke, K. *et al.* (2003). *Appl. Phys. Lett.,* **83**(14), 2970-2972.

Seibert, M. M., Ekeberg, T., Maia, F. R., Svenda, M., Andreasson, J., Josson, O., Odic, D., Iwan, B., Rocker, A., Westphal, D. *et al.* (2011). *Nature,* **470**(7332), 78-81.

Tiedtke, K., Feldhaus, J., Hahn, U., Jastrow, U., Nunez, T., Tschentscher, T., Bobashev, S., Sorokin, A., Hastings, J., Moller, S. *et al.* (2008). *J. Appl. Phy.*, **103**(9), 094511.

Yang, B., Wu, J., Raubenheimer, T. O. & Feng, Y. (2017). *J. Synchrotron Rad.,* **24**(3), 547.

Young, L., Kanter, E., Krassig, B., Li, Y., March, A., Pratt, S., Santra, R., Southworth, S., Rohringer, N., DiMauro, L. *et al*. (2010). *Nature,* **466**(7302), 56-61.

Zhu, D., Cammarata, M., Feldkamp, J. M., Fritz, D. M., Hastings, J. B., Lee, S., Lemke, H. T., Robert, A., Turner, J. L. & Feng, Y. (2012). *Appl. Phys. Lett.*, **101**, 034103.